\documentclass[prl,twocolumn,twoside,preprintnumbers,superscriptaddress,nofootinbib]{revtex4}

\usepackage{color}
\usepackage{amsmath}
\usepackage{graphicx}
\usepackage{dcolumn}
\usepackage{bm}
\usepackage[hyperfootnotes=false]{hyperref}
\usepackage{xspace}

\usepackage{textcomp}

\newcommand{\be}{\begin{equation}}
\newcommand{\ee}{\end{equation}}

\newcommand{\bea}{\begin{eqnarray}}
\newcommand{\eea}{\end{eqnarray}}

\newcommand{ \mysmall}[1]{\scriptscriptstyle #1} 

\newcommand{\ba} {\begin{eqnarray}}
\newcommand{\ea} {\end{eqnarray}}

\newcommand{\nn}{\nonumber}

\newcommand{\arXhref}[1]{\href{http://arxiv.org/abs/#1}{#1}}

\def\article{\@ifnextchar[{\earticle}{\oarticle}}

\begin{document}

\title{Revisiting Lepton Flavour Universality in B Decays}
\author{Ferruccio Feruglio}
\affiliation{Dipartimento di Fisica e Astronomia `G. Galilei', Universit\`a di Padova, Italy}
\affiliation{Istituto Nazionale Fisica Nucleare, Sezione di Padova, I--35131 Padova, Italy}
\author{Paride Paradisi}
\affiliation{Dipartimento di Fisica e Astronomia `G. Galilei', Universit\`a di Padova, Italy}
\affiliation{Istituto Nazionale Fisica Nucleare, Sezione di Padova, I--35131 Padova, Italy}
\author{Andrea Pattori}
\affiliation{Dipartimento di Fisica e Astronomia `G. Galilei', Universit\`a di Padova, Italy}
\affiliation{Physik-Institut, Universit\"at Z\"urich, CH-8057 Z\"urich, Switzerland}

\begin{abstract}
Lepton flavour universality (LFU) in B-decays is revisited by considering a class of semileptonic operators defined at a scale
$\Lambda$ above the electroweak scale $v$. The importance of quantum effects, so far neglected in the literature, is emphasised.
We construct the low-energy effective Lagrangian taking into account the running effects from $\Lambda$ down to $v$ through the one-loop 
renormalization group equations (RGE) in the limit of exact electroweak symmetry and QED RGEs from $v$ down to the $1 \,{\rm GeV}$ scale.
The most important quantum effects turn out to be the modification of the leptonic couplings of the vector boson $Z$ and the generation 
of a purely leptonic effective Lagrangian. Large LFU breaking effects in $Z$ and $\tau$ decays and visible lepton flavour violating (LFV) 
effects in the processes $\tau\to \mu\ell\ell$, $\tau\to\mu \rho$, $\tau\to\mu \pi$ and $\tau\to\mu \eta^{(\prime)}$ are induced.
\end{abstract}
\maketitle
%
{\bf Introduction}
Lepton flavour universality (LFU) tests are among the most powerful probes of the Standard Model (SM) and, 
in turn, of New Physics (NP) effects. In recent years, experimental data in $B$ physics hinted at deviations 
from the SM expectations, both in charged-current as well as neutral-current transitions. 
The statistically most significant data are: 
\begin{itemize}
\item 
An overall  $3.9 \sigma$ violation from the $\tau/\ell$ universality $(\ell = \mu,e)$ in the charged-current $b\to c$ decays~\cite{Lees:2013uzd,
Huschle:2015rga,Aaij:2015yra,Fajfer:2012vx}:
\begin{align}
R^{\tau/\ell}_{D^{(*)}}&=\frac{\mathcal{B}(\bar{B}\rightarrow D^{(*)}\tau\bar{\nu})_{\rm exp}/\mathcal{B}(\bar{B}\rightarrow D^{(*)}\tau\bar{\nu})_{\rm \mysmall SM}
}{\mathcal{B}(\bar{B}\rightarrow D^{(*)} \ell\bar{\nu})_{\rm exp}/\mathcal{B}(\bar{B}\rightarrow D^{(*)} \ell\bar{\nu})_{\rm \mysmall SM}},
\\
R_D^{\tau/\ell} &= 1.37\pm 0.17,  \quad \quad
R_{D^*}^{\tau/\ell} =  1.28\pm 0.08\,.
\label{R_D}
\end{align}
\item 
 A $2.6 \sigma$ deviation from $\mu/e$ universality in the neutral-current $b\rightarrow s$ transition~\cite{Aaij:2014ora}: 
\begin{equation}
R_K^{\mu/e} = \frac{\mathcal{B}(B\rightarrow K\mu^+\mu^-)_{\rm exp}}{\mathcal{B}(B\rightarrow Ke^+e^-)_{\rm exp}} =0.745^{+0.090}_{-0.074}\pm0.036\,,
\label{R_K}
\end{equation}
while $(R_K^{\mu/e})_{SM} =1$ up to few $\%$ corrections~\cite{Bordone:2016gaq}.
\end{itemize}
As argued in~\cite{Hiller:2014yaa,Hurth:2014vma,Altmannshofer:2014rta,Descotes-Genon:2015uva} by means of global-fit analyses, the explanation of the 
$R_K^{\mu/e}$ anomaly favours an effective 4-fermion operator involving left-handed currents, $(\bar{s}_L\gamma_\mu b_L)(\bar{\mu}_L\gamma_\mu \mu_L)$.
This naturally suggests to account also for the charged-current anomaly through a left-handed operator 
$(\bar{c}_L\gamma_\mu b_L)(\bar{\tau}_L\gamma_\mu \nu_L)$ which is related to $(\bar{s}_L\gamma_\mu b_L)(\bar{\mu}_L\gamma_\mu \mu_L)$ 
by the $SU(2)_L$ gauge symmetry~\cite{Bhattacharya:2014wla}. 
Clearly, this picture might work only provided NP couples much more strongly to the third generation than to the first two. 
Such a requirement can be naturally accomplished in two ways: i) assuming that NP is coupled, in the interaction basis, only to the third 
generation of quarks and leptons -- couplings to lighter generations are then generated by the misalignment between the mass and the 
interaction bases through small flavour mixing angles~\cite{Glashow:2014iga} -- and ii) if NP couples to different fermion generations 
proportionally to their mass squared~\cite{Alonso:2015sja}.
In the scenario i) LFU violation necessarily implies lepton flavour violating (LFV) phenomena. The same is not true in scenario ii) if the 
lepton family numbers are preserved.

In this work, we revisit the LFU in B-decays focusing on a class of semileptonic operators defined above the electroweak scale $v$
and invariant under the full SM gauge group, along the lines of Refs.~\cite{Fajfer:2012jt,Alonso:2014csa,Glashow:2014iga,Bhattacharya:2014wla,Buras:2014fpa,Alonso:2015sja,Calibbi:2015kma}. 
The main new development of our study is the construction of the low-energy effective Lagrangian taking into account the running of the 
Wilson coefficients of a suitable operator basis and the matching conditions when mass thresholds are crossed. 
The running effects from the NP scale $\Lambda$ down to the electroweak scale are included through the one-loop renormalization group equations (RGE) 
in the limit of exact electroweak symmetry~\cite{Jenkins:2013wua}. From the electroweak scale down to the $1 \,{\rm GeV}$ scale we use the QED RGEs.
By explicit calculations, we have checked that the scale dependence of the RGE contributions from gauge and top yukawa interactions cancels 
with that of the matrix elements in the relevant physical amplitudes. Such a program has not been carried out in the literature so far and it 
has significant implications on the conclusions of Refs.~\cite{Fajfer:2012jt,Alonso:2014csa,Glashow:2014iga,Bhattacharya:2014wla,Buras:2014fpa,Alonso:2015sja,Calibbi:2015kma}. 
The most important quantum effects turn out to be the modification of the leptonic couplings of the vector boson $Z$ and the generation 
of a purely leptonic effective Lagrangian. As a result, large LFV and LFU breaking effects in $Z$ and $\tau$ decays are induced.
\\

{\bf Effective Lagrangians}
%
If the NP contributions originate at a scale $\Lambda \gg v$, in the energy window above $v$ and below $\Lambda$
the NP effects can be described by an effective Lagrangian ${\cal L}\!=\!{\cal L}_{\rm SM}+{\cal L}_{\rm NP}$ invariant 
under the SM gauge group. Here we assume that NP is dominated by
\begin{align}
{\cal L}_{\rm \small NP}= ~
&\frac{C_1}{\Lambda^2} \left(\bar q_{3L}\gamma^\mu q_{3L}\right)\left(\bar \ell_{3L}\gamma_\mu \ell_{3L}\right) +
\nonumber\\
&\frac{C_3}{\Lambda^2} \left(\bar q_{3L}\gamma^\mu \tau^a q_{3L}\right)\left(\bar \ell_{3L}\gamma_\mu \tau^a \ell_{3L}\right).
\end{align}
We move from the interaction to the mass basis through the unitary transformations
\begin{align}
u_L \to V_u u_L && d_L \to V_d d_L &&V_u^\dagger V_d=V \,,\\
\nu_L \to U_e \nu_L && e_L \to U_e e_L \,, &&
\end{align}
where $V$ is the CKM matrix and neutrino masses have been neglected. We get
\begin{align}
\!\!\!{\cal L}_{\rm NP} \!=\!
\frac{1}{\Lambda^2}
[
&(C_1 \!+\! C_3) \,\lambda^u_{ij}\lambda^e_{kl}\, (\bar u_{Li}\gamma^\mu u_{Lj}) (\bar\nu_{Lk}\gamma_\mu \nu_{Ll})~+\nn\\
& (C_1 \!-\! C_3)\,\lambda^u_{ij}\lambda^e_{kl}\,\bar (u_{Li} \gamma^\mu u_{Lj}) (\bar e_{Lk}\gamma_\mu e_{Ll})~+\label{lag1}\nn\\
&(C_1 \!-\! C_3)\,\lambda^d_{ij}\lambda^e_{kl}\,(\bar d_{Li} \gamma^\mu d_{Lj})(\bar\nu_{Lk}\gamma_\mu \nu_{Ll})~+\nn\\
& (C_1 \!+\! C_3)\,\lambda^d_{ij}\lambda^e_{kl}\,(\bar d_{Li} \gamma^\mu d_{Lj})(\bar e_{Lk}\gamma_\mu e_{Ll})~+\nn\\
&2 C_3\!\left(\lambda^{ud}_{ij}\lambda^{e}_{kl}\,(\bar u_{Li} \gamma^\mu d_{Lj})(\bar e_{Lk}\gamma_\mu \nu_{Ll})\!+\! h.c.\right)
],
\end{align}
where
\begin{align}
\lambda^q_{ij}=V_{q3i}^*V_{q3j}~~~~\lambda^e_{ij}=U_{e3i}^*U_{e3j}~~~~\lambda^{ud}_{ij}=V_{u3i}^*V_{d3j}\,,
\end{align}
with $q=u,d$. These matrices are redundant since they satisfy the relations $\lambda^u=V \lambda^d V^\dagger$ and $\lambda^{ud}=V\lambda^d$.
We also observe that $\lambda^{f}$ are hermitian rank-1 matrices, satisfying $\lambda^f\lambda^f=\lambda^f$ and ${\rm tr}\lambda^f=1$. 
In summary, the free parameters of our Lagrangian are the ratios $(C_{1,3})/\Lambda^2$ and the two matrices $\lambda^d$ and $\lambda^e$.

Starting from the effective Lagrangian ${\cal L}_{\rm NP}$ at the scale $\Lambda$, at lower energies an effective Lagrangian is induced 
by RGE and by integrating out the heavy degrees of freedom. We will detail this procedure elsewhere. 
Here we summarize our results, obtained in a leading logarithmic approximation.

The effective Lagrangian describing the semileptonic processes $b \to s\ell\ell$ and $b \to s\nu\nu$ is~\cite{Buchalla:1995vs}
\begin{equation}
\label{eq:Heff}
  {\cal{L}}^{\mysmall \rm NC}_{\rm eff}=
  \frac{4 G_F}{\sqrt{2}}  \lambda_{bs}  \left( C^{ij}_\nu  \mathcal{O}^{ij}_\nu  +  C^{ij}_9  \mathcal{O}^{ij}_9  +  C^{ij}_{10}  \mathcal{O}^{ij}_{10}  \right)+h.c.\,,
\end{equation}
where $\lambda_{bs} \!=\! V_{tb}^{} V_{ts}^\ast$ and the operators $\mathcal{O}_{\nu}$ and $\mathcal{O}_{9,10}$ read
\begin{align}
  \mathcal{O}^{ij}_{\nu}  &= 
    \frac{e^2}{(4 \pi)^2} (\bar{s}_L \gamma_\mu b_L)(\bar{\nu}_i \gamma^\mu (1\!-\!\gamma_5) \nu_j)\,,\\
  \mathcal{O}^{ij}_{9} & = 
    \frac{e^2}{(4 \pi)^2} (\bar{s}_L \gamma_\mu b_L)(\bar{e}_i \gamma^\mu e_j)\,,\\
  \mathcal{O}^{ij}_{10}  &= 
    \frac{e^2}{(4 \pi)^2} (\bar{s}_L \gamma_\mu b_L)(\bar{e}_i \gamma^\mu \gamma_5 e_j)\,.
\end{align}
By matching ${\cal{L}}^{\mysmall \rm NC}_{\rm eff}$ with ${\cal L}_{\rm NP}$, we obtain:
\begin{align}
C^{ij}_9 = &-C^{ij}_{10} = \frac{4\pi^2}{e^2\lambda_{bs}}\frac{v^2}{\Lambda^2}
\,(C_{1} \!+\! C_{3})\,
\lambda^d_{23}\lambda^e_{ij} + \cdots\,,\\ 
C^{ij}_\nu = &\frac{4\pi^2}{e^2\lambda_{bs}}\frac{v^2}{\Lambda^2}
\,(C_{1} \!-\! C_{3})\,
\lambda^d_{23}\lambda^e_{ij} + \cdots \,,
\end{align}
where dots stand for RGE induced terms which are always subdominant, unless $C_1 = -C_3$ or $C_1 = C_3$. 
The latter condition, which can be realised in scenarios with vector leptoquark mediators~\cite{Calibbi:2015kma}, 
received a lot of attention in the literature as it allows to avoid the $B \!\to\! K^{(*)}\nu\bar\nu$ constraint.
We point out that such condition is not stable under quantum corrections.
RGE effects driven by the gauge interactions generate a rather large correction to $c_- = C_1-C_3$ 
at the electroweak scale
\begin{align}
\delta c_- \approx  - 0.03 \, C_3 \, \log\left( \!\frac{\Lambda}{m_Z} \!\right) \,,
\end{align}
which is of order $|\delta c_-| \sim 0.1$ for $C_3 = 1$ and $\Lambda \sim {\rm TeV}$.

The effective Lagrangian relevant for charged-current processes like $b\to c\ell\nu$ is given by
\begin{equation}
\label{eq:Heff}
\!\!\!{\cal{L}}^{\mysmall \rm CC}_{\rm eff} \!=\! -\frac{4 G_F}{\sqrt{2}} \, V_{cb} \, (C^{cb}_L)_{ij}\left( \bar c_L \gamma_\mu b_L  \right) \left( \bar e_{Li} \gamma^\mu \nu_{Lj}  \right) + h.c.\,,
\end{equation}
where the coefficient $(C^{cb}_L)_{ij}$ reads
\begin{equation}
(C^{cb}_L)_{ij} = \delta_{ij} - \frac{v^2}{\Lambda^2} \frac{\lambda^{ud}_{23}}{V_{cb}} \,C_3\, \lambda^{e}_{ij} 
\,.
\end{equation}

One of the effects due to ${\cal L}_{\rm NP}$ is the modification of the leptonic couplings of the vector bosons 
$W$ and $Z$. Focusing on the $Z$ couplings, which are the most tightly constrained by the experimental data, 
we find that
\begin{equation}
\!\!\mathcal{L}_Z = 
\frac{g_2}{c_W} \bar e_i \! \left( Z\!\!\!\!/ \,g^{ij}_{\ell L} P_L + Z\!\!\!\!/ \, g^{ij}_{\ell R} P_R \right) \! e_j + \frac{g_2}{c_W} \bar\nu_{Li} \, Z\!\!\!\!/ \,\, g^{ij}_{\nu L} \, \nu_{Lj} \,,
\end{equation}
where $g_{f L,R}= g_{f L,R}^{\mysmall \rm SM} + \Delta g_{f L,R}$, $c_W=\cos\theta_W$ and 
\begin{align}
\!\!\!\!\!
\Delta g^{ij}_{\ell L} &\!\simeq\! \frac{v^2}{\Lambda^2} \!
\left(\! 3y_t^2 c_- 
\lambda^{u}_{33} 
L_t \!+\! g_2^2 C_3 L_z \!+\! \frac{g_1^2}{3} C_1 L_z \!
\right)\!
\frac{\lambda^e_{ij}}{16\pi^2},
\\
\!\!\!\!\!
\Delta g^{ij}_{\nu L} &\!\simeq\! \frac{v^2}{\Lambda^2} \!
\left(\! 3y_t^2 c_+
\lambda^{u}_{33} L_t \!-\! g_2^2 C_3 L_z \!+\! \frac{g_1^2}{3} C_1 L_z \!
\right)\!
\frac{\lambda^e_{ij}}{16\pi^2},
\end{align}
with $L_{t} = \log\left(\Lambda/m_{t}\right)$, $L_{z} = \log\left(\Lambda/m_{Z}\right)$ and $\Delta g_{\ell R}=0$. 
The above expressions provide a good approximation of the exact results, which will be given elsewhere and  which have 
been obtained adding to the RGE contributions from gauge and top yukawa interactions the explicit one-loop matrix element 
with the $Z$ four-momentum set on the mass-shell. The scale dependence of the RGE contribution cancels with that 
of the matrix element dominated by a quark loop. Hereafter, we systematically neglect corrections of order 
$m_q^2/(16\pi^2\Lambda^2)$ when $q=u,d,c,s,b$. 

Quantum effects generate also a purely leptonic effective Lagrangian, as well as corrections to the semileptonic interactions.
After running the Wilson coefficients from $\Lambda$ down to the electroweak scale and integrating out the $W$, $Z$ and the 
heavy quarks $c$, $b$, and $t$, we get the leading terms:
\begin{align}
\!{\cal L}^\ell_{\rm eff} &\!=\!-\frac{4G_F}{\sqrt{2}}\lambda^e_{ij}
\bigg[
(\overline{e}_{Li}\gamma_\mu e_{Lj})
{\sum}_\psi 
\overline{\psi}\gamma^\mu \psi \left( 2 g^{\mysmall\rm Z}_\psi c^e_t - Q_\psi c^e_\gamma \right)
\nn\\
&\!\!\!\!\!\!\!\!\!\!\!+
c^{cc}_t (\overline{e}_{Li}\gamma_\mu \nu_{Lj}) 
(\overline{\nu}_{Lk} \gamma^\mu e_{Lk} + \overline{u}_{Lk} \gamma^\mu V_{kl} d_{Ll} )\!+h.c.
\!\bigg],
\label{eq:4-fermion}
\end{align}
where $\psi=\{\nu_{Lk},e_{Lk,Rk},u_{L,R},d_{L,R},s_{L,R}\}$ and $g^{\mysmall\rm Z}_\psi$ is the fermionic $Z$ coupling 
defined as $g^{\mysmall\rm Z}_\psi = T_3(\psi) -Q_\psi \sin^{2}\theta_W$. 
In eq.~(\ref{eq:4-fermion}) we neglected additional neutrino interactions, irrelevant in our analysis.
Finally, the coefficients $c^{e,cc}_t$ and $c^{e}_\gamma$ are given by
\begin{align}
\!\!\!\!\! c^e_t & \!= \frac{3y^2_t}{32\pi^2} \frac{v^2}{\Lambda^2}(C_1\!-\!C_3) \,\lambda^u_{33} \, \log\frac{\Lambda^2}{m_t^2}\,, 
\nn\\
\!\!\!\! c^{cc}_t & \!= \frac{3 y^2_t}{16\pi^2} \,\frac{v^2}{\Lambda^2}~ C_3 \,\lambda^u_{33} \left[ \log\frac{\Lambda^2}{m_t^2}+\frac{1}{2}\right]\,, 
\nn\\
\!\!\!\!\! c^e_\gamma &\!=\! \frac{e^2}{48\pi^2} \! \frac{v^2}{\Lambda^2} 
\bigg[
(3C_3 \!-\! C_1)\log\frac{\Lambda^2}{\mu^2} - (C_1 \!+\! C_3)\lambda^d_{33}\log\frac{m_b^2}{\mu^2} 
\nn\\
&\qquad~~~ +2(C_1\!-\!C_3)\!\left(\!\lambda^u_{33}\log\frac{m_t^2}{\mu^2} \!+ \lambda^u_{22}\log\frac{m_c^2}{\mu^2}\right)\!
\bigg].
\label{eq:L^ell_eff}
\end{align}
The residual scale dependence is removed by evaluating the matrix elements in the low energy theory, which includes the light quarks
$u$, $d$, $s$. For simplicity, we have done this within the quark model, by assuming for $u$, $d$ and $s$ a common constituent mass 
$\mu\approx 1$ GeV.

As shown by eq.~(\ref{eq:L^ell_eff}), ${\cal L}^\ell_{\rm eff}$ receives one-loop induced RGE contributions of order $y_t^2/16\pi^2$ and $e^2/16\pi^2$. 
The former arises from the top-quark yukawa interactions and affects both the neutral and charged currents.  
On the contrary, the effects induced by the SM gauge interactions cancel completely in the charged current and only partially in the neutral current,
where they are proportional to $e^2$ and to the electromagnetic current. \\

{\bf Observables}
We proceed by analysing the phenomenological implications of our low-energy theory.
We will revisit first the anomalies in the processes $B\to K\ell\bar\ell$ and $B\to D^{(*)}\ell\bar\nu$ 
under the constraints imposed by $B\to K \nu\bar\nu$. Then, we will study observables receiving contributions at the loop-level, 
so far overlooked in the literature, which include both LFV and LFU breaking effects in $Z$ and $\tau$ decays.

In our model, $R^{\mu /e}_{K}$ is approximated by the expression
\begin{equation}
R^{\mu /e}_{K} \approx  
\frac{|C_9^{\mu\mu} + C_9^{\mysmall\rm SM}|^2}{| C_9^{ee} + C_9^{\mysmall\rm SM}|^2}\,,
\label{eq:R_K}
\end{equation}
where $C^{\mysmall\rm SM}_9 \!\approx\! 4.2$.  
The experimental central value $R^{\mu /e}_{K} \!\approx\! 0.75$ is reproduced for $C_9^{\mu\mu} \!\approx\! -0.5$ if we assume $C_{9}^{ee} \!=\! 0$. 
In particular, we find that
\begin{align}
R^{\mu /e}_{K} & \approx  1 - 0.28 \, \frac{(C_1 + C_3)}{\Lambda^2({\rm TeV})} \frac{\lambda^d_{23}\,\lambda^e_{22}}{10^{-3}}\,.
\label{eq:R_K_num}
\end{align}

The expression for $R^{\tau/\ell}_{D^{(*)}}$ reads
\begin{equation}
R^{\tau/\ell}_{D^{(*)}} = \frac{\sum_{j} |(C^{cb}_{L})_{3j}|^2}{\sum_{j} |(C^{cb}_{L})_{\ell j}|^2}\,,
\label{eq:R_D}
\end{equation}
where $\ell =e,\mu$. Assuming that $\lambda^{e}_{22} \ll \lambda^{e}_{33}\sim 1$, we find 
\begin{align}
R^{\tau/\ell}_{D^{(*)}} &\approx 1 - \frac{0.12 \, C_3}{\Lambda^2({\rm TeV})} \left( \lambda^d_{33} + \frac{V_{cs}}{V_{cb}}\lambda^d_{23} \right) 
\,.
\label{eq:R_D_num}
\end{align}
The condition $\lambda^{e}_{22} \!\ll\! \lambda^{e}_{33}$ is justified by the non observation of LFU breaking effects 
in the $\mu/e$ sector up to the $\lesssim 2\%$ level~\cite{Agashe:2014kda,Greljo:2015mma}, leading to the upper bound 
$\lambda^e_{22} \lesssim 0.1$ once the anomaly in the $\tau/\ell$ sector is explained. 
In our estimates we always set $\lambda^d_{11}=0$, as well as $\lambda^e_{11}=0$ which implies $\lambda^e_{22}\sim (\lambda^{e}_{23})^2$.

As already noted in~\cite{Calibbi:2015kma}, non trivial constraints arise from the process $B\to K \nu\bar\nu$.
Defining $R^{\nu\nu}_{K}$ as
\begin{equation}
R^{\nu\nu}_{K} = \frac{\mathcal{B}(B\to K \nu\bar\nu)}{\mathcal{B}(B\to K \nu\bar\nu)_{\mysmall\rm SM}}
= \frac{\sum_{ij} |C^{\mysmall \rm SM}_\nu \,\delta^{ij}+ C^{ij}_\nu|^2}{3|C^{\mysmall \rm SM}_\nu |^2}\,,
\end{equation}
where $C^{\mysmall \rm SM}_\nu \approx -6.4$ and exploiting the properties ${\rm tr}\lambda^f = 1$ and  $\sum_{ij} |\lambda^f_{ij}|^2 = 1$,
we obtain
\begin{equation}
R^{\nu\nu}_{K} \approx 1 + \frac{0.6 \,c_-}{\Lambda^2({\rm TeV})} 
\left(\frac{\lambda^d_{23}}{0.01} \right)
+ \frac{0.3 \, c^2_- }{\Lambda^4({\rm TeV})} 
\left(\frac{\lambda^d_{23}}{0.01} \right)^{\!2},
\end{equation}
while the experimental bound reads $R^{\nu\nu}_{K} <  4.3$~\cite{Lees:2013kla}.
If LFU effects arise from LFV sources, LFV phenomena are unavoidable~\cite{Glashow:2014iga}. 
In our setting, it turns out that~\cite{Crivellin:2015era}
\begin{align}
\!\!\mathcal{B}(B \!\to\! K \tau\mu) & \approx 4 \!\times\! 10^{-8} \left| C_{9}^{\mu\tau}\right|^2 
\!\approx\! 10^{-7} 
\left| \frac{C_{9}^{\mu\mu}}{0.5}\, \frac{0.3}{\lambda^e_{23}}\right|^2,
\end{align}
where we have exploited the relation $C_{9}^{\mu\mu}/C_{9}^{\mu\tau} \!\approx \! \lambda^e_{23}$
and set $|C_{9}^{\mu\mu}| \approx 0.5$ to accommodate the $R^{e/\mu}_K$ anomaly.
The above prediction is orders of magnitude below the current bound $\mathcal{B}(B\to K \tau\mu) \leq 4.8 \times 10^{-5}$~\cite{Amhis:2014hma}.

Modifications of the leptonic $Z$ couplings are constrained by the LEP measurements
of the $Z$ decay widths, left-right and forward-backward asymmetries.
The bounds on lepton non-universal couplings are~\cite{Agashe:2014kda}
\begin{align}
\frac{v_\tau}{v_e} = 0.959\; (29)\,, \qquad
\frac{a_\tau}{a_e} = 1.0019\; (15)\,,
\label{eq:zpole_pdg}
\end{align}
where $v_\ell$ and $a_\ell$ are the vector and axial-vector couplings, respectively, defined as 
$v_\ell = g^{\ell\ell}_{\ell L} + g^{\ell\ell}_{\ell R}$ and 
$a_\ell = g^{\ell\ell}_{\ell L} - g^{\ell\ell}_{\ell R}$. 
We get
\begin{equation}
\frac{v_\tau}{v_e} \simeq 1 - \frac{2\,\Delta g^{33}_{\ell L}}{(1-4s^2_W)}\,,\qquad
\frac{a_\tau}{a_e} \simeq 1 - 2\,\Delta g^{33}_{\ell L}
\,,
\label{eq:zpole_th}
\end{equation}
leading to the following numerical estimates
\begin{align}
\frac{v_\tau}{v_e} & \approx 1 - 0.05 \, \frac{\left( c_- + 0.2 \,C_3 \right)}{\Lambda^2({\rm TeV})}\,, 
\nonumber\\
\frac{a_\tau}{a_e} & \approx 1 - 0.004 \, \frac{\left( c_- + 0.2 \,C_3 \right)}{\Lambda^2({\rm TeV})}\,,
\label{eq:zpole_num}
\end{align}
where we took $\lambda^u_{33}\sim \lambda^e_{33}\sim 1$ and, hereafter, we set $\Lambda = 1$ TeV in the argument of the logarithm.
Moreover, modifications of the $Z$ couplings to neutrinos affect the extraction of the number 
of neutrinos $N_\nu$ from the invisible Z decay width. We find that
\begin{align}
N_\nu = 2 + \, \left(\frac{g^{33}_{\nu L}}{g^{\mysmall \rm SM}_{\nu L}}\right)^2 \simeq\, 3 + 4 \, \Delta g^{33}_{\nu L}\,,
\end{align}
leading to the following numerical estimate
\begin{equation}
N_\nu \approx 3 + 0.008 \, \frac{\left( c_+ - 0.2 \,C_3 \right)}{\Lambda^2({\rm TeV})}\,,
\end{equation}
to be compared with the experimental result~\cite{Agashe:2014kda}
\begin{equation}
N_\nu = 2.9840 \pm 0.0082\,.
\end{equation}
Finally, we have checked that $\mathcal{B}(Z \to \mu^\pm \tau^\mp)$ is always well below the current experimental bound.

LFU breaking effects in $\tau \to \ell \bar\nu \nu$ (with $\ell_{1,2}=e,\mu$) are described by the observables
\begin{align}
\!\!\!\!\!R^{\tau/\ell_{1,2}}_\tau = \frac{\mathcal{B}(\tau \to \ell_{2,1} \nu\bar\nu)_{\rm exp}/\mathcal{B}(\tau \to \ell_{2,1} \nu\bar\nu)_{\rm SM}}{\mathcal{B}(\mu \to e \nu\bar\nu)_{\rm exp}/\mathcal{B}(\mu \to e \nu\bar\nu)_{\rm SM}} \,,
\end{align}
and are experimentally tested at the few \textperthousand~level~\cite{Pich:2013lsa}
\begin{align}
\!\!\!R^{\tau/\mu}_\tau = 1.0022 \pm 0.0030 \,,~~ R^{\tau/e}_\tau = 1.0060 \pm 0.0030 \,.
\label{eq:tau_LFU_data}
\end{align}
We find
\begin{align}
R^{\tau/\ell}_\tau &\simeq 1 + 2\,c^{cc}_t \lambda^{e}_{33}
\approx 1 + \frac{0.008 \, C_3}{\Lambda^2({\rm TeV})}
\label{eq:tau_LFU}
\,.
\end{align}
\begin{figure}[ht!]
\centering
\includegraphics[width=0.85\linewidth]{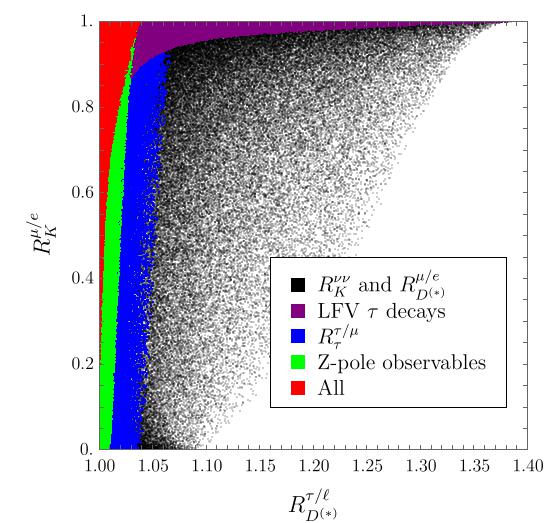}
\\ ~ \\
\includegraphics[width=0.85\linewidth]{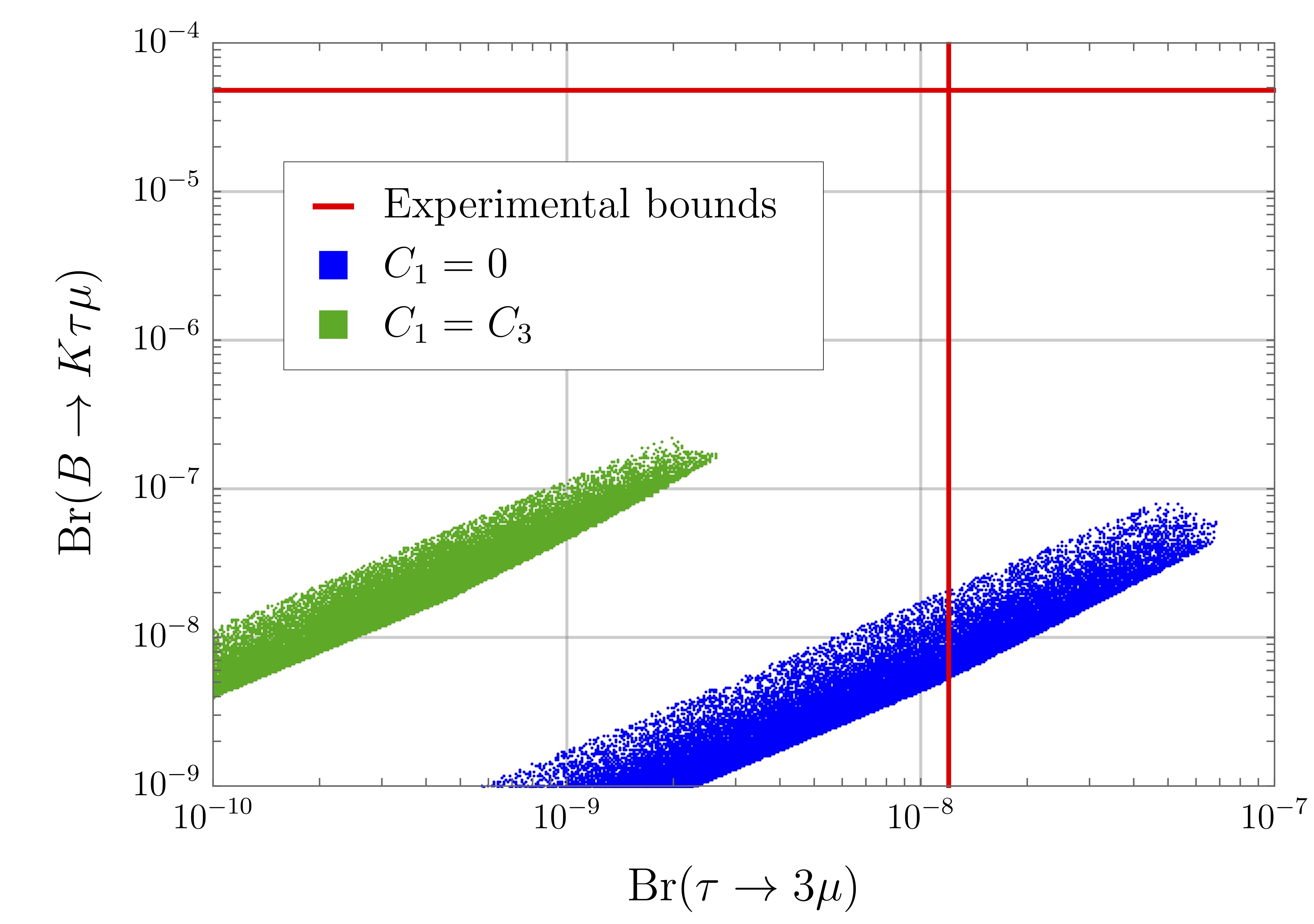}
\vspace{-10pt}
\caption{Upper plot: $R^{\mu/e}_{K}$ vs.~$R^{\tau/\ell}_{D^{(*)}}$ for $C_1 =0$, $|C_3| \leq 3$, $|\lambda^d_{23}|\leq 0.04$ and $|\lambda^e_{23}|\leq 1/2$. 
The allowed regions are coloured according to the legend.
Lower plot: $\mathcal{B}(B \to K \tau\mu)$ vs. $\mathcal{B}(\tau\to 3\mu)$ for $|\lambda^d_{23}|=0.01$, $C_1=C_3$ (green points) or $C_1=0$ (blue points)
imposing all the experimental bounds except $R^{\tau/\ell}_{D^{(*)}}$.}
\label{fig1}
\end{figure}
The effective Lagrangian of eq.~(\ref{eq:4-fermion}) generates LFV processes such as $\tau \!\to\! \mu\ell\ell$ and $\tau \!\to\! \mu P$ with $P \!=\! \pi,\eta,\eta^\prime,\rho$, etc.  
The most sensitive channels, taking into account their NP sensitivities and experimental resolutions, are $\tau\to \mu\ell\ell$, $\tau\to \mu\rho$ and $\tau\to \mu\pi$.
For $\tau\to \mu\ell\ell$ we find
\begin{equation}
\!\!\!\!\!\frac{\mathcal{B}(\tau\to \mu\ell\ell)}{\mathcal{B}(\tau\to \mu\nu\bar\nu)} = 
|\lambda^e_{23}|^2
\left[
(1+\delta_{\ell\mu})\!\left(c_{\rm \mysmall LR} - c^e_t\right)^2 \!+ c_{\rm \mysmall LR} ^2
\right]\,,
\end{equation}
where $c_{\rm \mysmall LR} = 2s^2_{\mysmall W} \, c^e_t +c^e_\gamma$.
If $c_- \sim \mathcal{O}(1)$, we obtain
\begin{equation}
\mathcal{B}(\tau\to 3\mu) \approx 5\times 10^{-8} \,
\frac{c_{-}^{\,2} }{\Lambda^4({\rm TeV})} 
\left(\frac{\lambda^e_{23}}{0.3}\right)^2 \,,
\label{eq:tau3mu_num}
\end{equation}
while the current bound is $\mathcal{B}(\tau\to 3\mu) \leq 1.2 \times 10^{-8}$~\cite{Amhis:2014hma}.
Setting $c_-(\Lambda) =0$ leads to $\mathcal{B}(\tau\to 3\mu) \approx 4\times 10^{-9}$ for 
$\Lambda = 1$ TeV, $\lambda^e_{23}=0.3$ and $C_1=C_3=1$, yet within the future expected 
experimental sensitivity. Moreover, it turns out that $1.5 \lesssim \mathcal{B}(\tau \!\to\! 3\mu)/\mathcal{B}(\tau \!\to\! \mu ee)\lesssim 2$.
Finally, employing the general formulae of ref. \cite{Brignole:2004ah}, we find
\begin{align}
\!\!\!\!\!\!\mathcal{B}(\tau\to\mu\rho) 
&\approx 
2\,|\lambda^e_{23}|^2 \left[(2s^2_{\mysmall W}-1) c^e_t + c^e_\gamma \right]^2 \mathcal{B}(\tau\to\nu\rho)
\nonumber\\
& \approx 5 \times 10^{-8} \frac{(c_- - 0.28C_3)^2}{\Lambda^4({\rm TeV})} 
\left(\frac{\lambda^e_{23}}{0.3}\right)^2
\,,
\label{eq:taumurho_num}
\end{align}
and
\begin{align}
\!\!\!\!\!\!\mathcal{B}(\tau\to\mu\pi) 
&\approx 
2\,|\lambda^e_{23}|^2 \left[ c^e_t \right]^2 \mathcal{B}(\tau\to\nu\pi)
\nonumber\\
& \approx 8 \times 10^{-8} \frac{c_- ^2}{\Lambda^4({\rm TeV})} 
\left(\frac{\lambda^e_{23}}{0.3}\right)^2
\,,
\label{eq:taumurho_num}
\end{align}
where the current bounds are $\mathcal{B}(\tau\to \!\mu\rho) \leq 1.5 \times 10^{-8}$
and $\mathcal{B}(\tau\to\mu\pi) \leq 2.7\times 10^{-8}$~\cite{Amhis:2014hma}.

We discuss now the necessary conditions to accommodate the B-physics anomalies and their phenomenological implications.
Two scenarios are envisaged: i) $C_1 = 0$ and $C_3 \neq 0$ and ii) $C_1 = C_3$. In both cases, the correct pattern of 
deviation from the SM expectations is reproduced for $C_3 < 0$, $|\lambda^d_{23}/V_{cb}|<1$ and $\lambda^d_{23}<0$, 
see eqs.~(\ref{eq:R_K_num}), (\ref{eq:R_D_num}). Moreover, for $|C_3| \sim \mathcal{O}(1)$, the upper bound $\Lambda \lesssim 1$ TeV 
and the lower bound $|\lambda^e_{23}| \gtrsim 0.1$ are also predicted. 
The major differences between the two scenarios concern the impact of the constraints from $Z$-pole and $\tau$ observables. 
In particular, from eqs. (\ref{eq:zpole_pdg}) and (\ref{eq:zpole_num}) we learn that NP effects in $v_\tau/v_e$ and $a_\tau/a_e$ 
are uncomfortably large in scenario i) while they are under control in ii). Similarly, $\mathcal{B}(\tau\to 3\mu)$ is one order of magnitude 
larger in i) than in ii), see eq.~(\ref{eq:tau3mu_num}) and following discussion. 
Most importantly, we find that $R^{\tau/\ell}_\tau$ strongly disfavours an explanation of the $R^{\tau/\ell}_{D^{(*)}}$ anomaly 
based on left-handed effective operators, see eqs.~(\ref{eq:R_D_num}), (\ref{eq:tau_LFU}).
This is confirmed by the upper plot of fig.~\ref{fig1} (where, to be conservative, we didn't impose the strong bound from $R^{\tau/e}_\tau$)
showing $R^{\mu/e}_{K}$ vs.~$R^{\tau/\ell}_{D^{(*)}}$ in the scenario i).
The overall picture doesn't change in scenario ii) as the $R^{\tau/\mu}_\tau$ bound is unchanged. 
In the lower plot of fig.~\ref{fig1}, we show $\mathcal{B}(B \to K \tau\mu)$ vs. $\mathcal{B}(\tau\to 3\mu)$. 
Considering the current and expected future experimental sensitivities, we conclude that $\tau\to 3\mu$ is a 
more powerful probe than $B\to K\tau\mu$ of both scenarios, especially i).
\\

{\bf Conclusions}
Recent experimental data hinting at non-standard LFU breaking effects in semileptonic $B$-decays~\cite{Lees:2013uzd,Huschle:2015rga,Aaij:2015yra,Aaij:2014ora} 
stimulated many theoretical investigations of NP scenarios~\cite{Hiller:2014yaa,Hurth:2014vma,Altmannshofer:2014rta,Descotes-Genon:2015uva,Fajfer:2012jt,Alonso:2014csa,Glashow:2014iga,Bhattacharya:2014wla,Buras:2014fpa,Alonso:2015sja,Calibbi:2015kma,Altmannshofer:2014cfa}.
In this work, we revisited LFU in B-decays assuming a class of gauge invariant
semileptonic operators at the NP scale $\Lambda \gg v$, as in Refs.~\cite{Fajfer:2012jt,Alonso:2014csa,Glashow:2014iga,Bhattacharya:2014wla,Buras:2014fpa,Alonso:2015sja,Calibbi:2015kma}.
We constructed the low-energy effective Lagrangian taking into account the running effects from $\Lambda$ down to $v$ through the one-loop RGEs in the limit of exact 
electroweak symmetry and QED RGEs from $v$ down to the $1 \,{\rm GeV}$ scale. 
At the quantum level, we find that the leptonic couplings of the $W$ and $Z$ vector bosons are modified. 
Moreover, quantum effects generate also a purely leptonic effective Lagrangian, as well as corrections to 
the semileptonic interactions. The main phenomenological implications of these findings are the generation of large 
LFU breaking effects in $Z$ and $\tau$ decays, which are correlated with the B-anomalies, and $\tau$ LFV processes. 
Overall, the experimental bounds on $Z$ and $\tau$ decays significantly constrain LFU breaking effects in B-decays, 
challenging an explanation of the current non-standard data~\cite{Lees:2013uzd,Huschle:2015rga,Aaij:2015yra,Aaij:2014ora},
at least in the framework adopted here. 
Interestingly, if LFU breaking effects arise from LFV sources, the most sensitive LFV channels are not $B$-decays, 
as commonly claimed in the literature but, instead, $\tau$ decays such as $\tau\to \mu\ell\ell$ and $\tau\to \mu \rho$.
Although our results have been obtained in the context of an effective Lagrangian dominated by left-handed operators,
the present work shows that electroweak radiative effects should be carefully analysed in any 
framework addressing the explanation of B-anomalies. 

{\bf Acknowledgements}
This work was supported in part by the MIUR-PRIN project 2010YJ2NYW and by the European Union network FP10 ITN ELUSIVES 
(H2020-MSCA-ITN-2015-674896) and INVISIBLES-PLUS (H2020- MSCA-RISE-2015-690575).
The research of P.P. is supported by the ERC Advanced Grant No.  267985 (DaMeSyFla), by the research grant TAsP, and by the INFN. 
The research of A.P. is supported by the Swiss National Science Foundation (SNF) under contract 200021-159720.

\newpage

\end{document}